\documentclass[amsmath,twocolumn,prl,floatfix]{revtex4}

\usepackage{color}
\usepackage{amssymb}
\usepackage{amsmath}
\usepackage{epsfig}
\usepackage{graphics}
\usepackage{graphicx}
\usepackage{dcolumn}% Align table columns on decimal point
\usepackage{subfigure}
\usepackage{bm}% bold math

\def\be{\begin{equation}}
\def\ee{\end{equation}}

\begin{document}

\title{Equations of Evolutionary Dynamics in High Dimensions}

\author{Alfred Ajay Aureate R.} 
\affiliation{The Institute of Mathematical Sciences,
CIT Campus, Taramani, Chennai 600113, India}
\author{Vaibhav Madhok}
\email{vmadhok@gmail.com}
\affiliation{Department of Physics, Indian Institute of Technology Madras, Chennai, India 600036}

\date{\today}

\begin{abstract}
We study quasi-species and closely related evolutionary dynamics like the replicator-mutator equation in high dimensions. 
In particular, we show that under certain conditions the fitness of almost all quasi-species becomes independent of mutational probabilities and the initial frequency distributions of the sequences in high dimensional sequence spaces. This result is the consequence of the concentration of measure on a high dimensional hypersphere and its extension to Lipschitz functions known %knows
as the Levy's Lemma. Therefore, evolutionary dynamics almost always yields the same value for fitness of the quasi-species, independent of the mutational process and initial conditions, and is quite robust to mutational changes and fluctuations in initial conditions. Our results naturally extend to any Lipschitz function whose input parameters are the frequencies of individual constituents of the quasi-species. This suggests that the functional capabilities of high dimensional quasi-species are robust to fluctuations in the mutational probabilities and initial conditions. We discuss the consequences of our study for the replicator-mutator equation. 
                           \end{abstract}

\pacs{PACS numbers}

\maketitle

\section{Introduction}

Living systems and life processes show a remarkable order despite the role of chance and mutational
processes underlying its origin.
Why are living systems so well adapted to their environment?   On the one hand, the performance of many biological systems, characterized as physical processes, is near optimal and close to the limits set by the laws of physics \cite{Bialek}. In the domain of biochemical processes, the  enzymes for example, serve as nearly optimal catalysts. On the other hand,  evolution viewed as a complex dynamical process with mutations, its essential fuel, being stochastic in nature, probably does not optimize anything. How does this near optimality and robustness in the presence of stochasticity arise?

The slogan, ``survival of the fittest" seems to be the accepted answer to the question posed above. However, without a proper definition of ``fitness", the question of adaptation remains contentious at best, and largely unanswered. Without a proper quantification of fitness, the above argument reduces to ``survival of the survivors",  which by its circular nature is an uncomfortable position to study evolutionary biology. 

One of the motivations behind the concept of quasi-species, introduced by Eigen and Schuster, was to be able to make precise statements about the notion of the survival of the fittest \cite{eigen77, eigen71}.  Quasi-Species is an ensemble with a well defined distribution of mutants that is a result of the evolutionary process involving selection and mutation. Selection acts on the quasi-species as a whole and the most optimal ensemble survives.
Quasi-Species sheds light on the role of chance in the process of adaptation by taking into account the role of errors in the process of replication which results in the generation of an ensemble of closely related species instead of a single fittest constituent. The equilibrium distribution resulting from the selection mutation process depends not only on the replication rates of individual constituents but also on the erroneous replication of the entire population.
Therefore, natural selection as an optimization is not directed toward the single fittest variant, but towards the ensemble which evolve to maximize its average replication rate. In general, the average replication rate, 
also known as the mean fitness will depend on the relative frequencies of the variants which in turn depends on the underlying mutational probabilities. Therefore, while the quasi-species formulation shows the role of chance in the process of adaptation, the near optimal adaptation observed is often attributed to the ``fitness" of the whole quasi-species.
After all, as mentioned above, living systems including biochemical processes like enzyme functions show efficient adaptation regardless of what role randomness might have had. Moreover, the mutational probabilities that cause cross-coupling between the individual variants have their origins in quantum mechanics and in general, should not be assumed to be fixed in the entire course of evolution.
Our work shows the robustness of the fitness function even when these assumptions are relaxed. This we do by mapping the solution of quasi-species equation to 
points on the surface of an $n$ dimensional hyper-sphere and invoking properties of concentration of measure as $n$ becomes large.
Therefore, for sufficiently large mutational rates and almost all initial conditions, the resulting quasi-species at equilibrium are equally fit and more importantly show quantitatively similar functional capabilities.

 To the best of our knowledge, ours is the first study that gives perspective on evolutionary dynamics from the point of view of high dimensional geometry. We certainly do not claim to have a solution to the near optimality and robustness of life processes, neither we claim that an answer is possible at all. However, our work does suggest that evolutionary dynamics can benefit from studies in statistical mechanics and high dimensional geometry. Indeed, application of maximum entropy methods and thermodynamics has found some success in addressing some  fundamental questions in biology \cite{england}. 

Our study can be extended to analyzing other kinds of evolutionary scenarios like the replicator-mutator equation. Here, unlike the quasi-species, we cannot talk about ``fitness" but nonetheless can still view the dynamics taking place on the surface of a hyper-sphere.
%Evolutionary dynamics of  

\section{Quasi-species equation}

%The quasi-species model is an attempt to address these questions.
%Quasispecies represents the evolution of high-mutation-rate viruses such as HIV and sometimes single genes or molecules within the genomes of other organisms (Ref. \cite{holland}\cite{domingo}\cite{wilke}). %Quasispecies models have also been proposed by Jose Fontanari and Emmanuel David Tannenbaum to model the evolution of sexual reproduction (Ref. \cite{tannenbaum}). %Quasispecies was also shown in compositional replicators (based on the Gard model for abiogenesis) (Ref. \cite{gross}) and was also suggested to be applicable to describe cell's replication, which amongst other things requires the maintenance and evolution of the internal composition of the parent and bud.

%Quasispecies describes quantitatively a simple information sequence evolution in terms of sequence length, population size, and mutation and selection intensities. This model can be used to characterize roughly the hypothetical prebiotic polynucleotide sequence evolution and to illustrate mathematically general features of biological evolution (Ref. \cite{redko}).

Quasi-species is an ensemble with a well defined distribution of mutants that is a result of the evolutionary process involving selection and mutation (Ref. \cite{quasispecies}). Selection acts on the quasi-species as a whole and the most optimal ensemble survives.

Quasi-species as an ensemble of related genotypes is given by

\begin{equation}
\frac{dX}{dt} = WX - f(X).X 
\label{dxdt}  \end{equation}

The vector X consists of the population densities of the individual sequences,

\begin{equation}
X = (x_{1},x_{2},...,x_{n}), 
\label{xs}  \end{equation}

The matrix $W$ consists of individual replication rates, $a_{i},i=1,2,...n$, along with the mutation rates for transition between individual sequences, $i$ and $j$, given by $Q_{ij}$.

\[
W = 
\begin{bmatrix}
    a_{1}Q_{11} & a_{2}Q_{12} & \dots  & a_{n}Q_{1n} \\
    a_{1}Q_{21} & a_{2}Q_{22} & \dots  & a_{n}Q_{2n} \\
    \dots & \dots & \dots & \dots \\
    a_{1}Q_{n1} & a_{2}Q_{n2} & \dots  & a_{n}Q_{nn}
\end{bmatrix}
\]

Here, if we consider only point mutations of nucleotides of length $m$ (where $n=4^{m}$), then for each row, we would have only $3m+1$ non-zero elements (including self-replication or non-replication of a base). For higher values of $m$, this matrix becomes very sparse.

The total size of the population is a constant if we have

\begin{equation}
    f(X) = \sum_{i=1}^{n} a_{i}x_{i}/\sum_{i}^{n} x_{i}
\label{fx}\end{equation}

The equilibrium of Eqn. (\ref{dxdt}) is given by solving the eigenvalue problem

\begin{equation}
 WX = \lambda X
\label{WX}  \end{equation}

The fact that the above system will have a unique largest positive eigenvalue is guaranteed by the Frobenius-Perron theorem \cite{perron07, frobenius12}.

We are more interested in the largest (positive) eigen value. The largest eigenvalue gives the average replication rate of the quasi-species, $\lambda_{max} = \sum_{i=1}^{n} a_{i}x_{i}/\sum_{i}^{n} x_{i}$ and the corresponding eigenvector gives the frequency distribution, $X_{eq} = (x_{1},x_{2},...,x_{n})$, at equilibrium.

The equilibrium frequency distribution, $X_{eq}=(x_{1},x_{2},\dots x_{n})$ could be normalized (as it represents probability) for simplicity, i.e. $\sum_{i=1}^{n} x_{i}=1$. Hence, $f(X)=\sum_{i=1}^{n} a_{i}x_{i}$, for normalized $X$. If we assume that initially the (non-normalized) $x_{i}$'s are independent and identically distributed (IID) and are picked from an exponential distribution with mean $\lambda=1$, then from $Appendix-1$, we could see that for very high values of $n$, the normalized $x_{i}$'s could also be assumed to be IID variables, picked from an exponential distribution with mean $1/n$.

\subsection{Using Levy's lemma for showing concentration around the mean}

Let us assume that the function $f:D\rightarrow  {\rm I\!R}$ is Lipschitz continuous with Lipschitz constant $\eta$ (with respect to the Euclidean norm), where $D=[0,1]^{n}/[0,\delta]^{n}$, meaning that at least one of the coordinates takes value more than $\delta\ll1$ (in order to avoid singularity around the neighbourhood of origin). We could then see that the square root of the normalized equilibrium frequency $\sqrt{X_{eq}}=(\sqrt{x_{1}},\sqrt{x_{2}},\dots,\sqrt{x_{n}})$ is almost uniformly distributed over an n-dimensional hypersphere (for higher values of $n$). Hence, we could modify Levy's lemma (from Ref. \cite{gerken}) for these points to show that,

\begin{equation}
    \text{Pr}\{ |f(X) - E[f(X)]| \geq \epsilon \} \leq \exp\left(-\frac{Kn\epsilon^{2}}{\eta^{2}}\right)
\label{levy}\end{equation}

for all $\epsilon \geq 0$, as explained in detail in \textit{Appendix-2}.

In fact, $f(X)$ from Eqn. (\ref{fx}) is indeed Lipschitz continuous with Lipschitz constant $\eta=\sqrt{n}a_{max}$, where $a_{max}=\max_{i}(a_{i})$, as shown in \textit{Appendix-3}.
Now, to improve the upper bound further, we could additionally assume $a_{i}$'s to satisfy the conditions, $0 \leq a_{1},a_{2},\dots ,a_{n} \leq a_{max}\leq C/n$ and $a_{1} + a_{2} + \dots a_{n} = 1$, where $C$ is some positive constant. This is very likely to be satisfied for higher values of $n$ (as implied by \textit{Appendix-1}). Then, from \textit{Appendix-3}, we see that,

\begin{equation}
\begin{split}
    |f(X)-f(Y)| &\leq \sqrt{n}a_{max}\left\Vert X-Y\right\Vert_{2}\\
    &\leq \frac{C}{\sqrt{n}}\left\Vert X-Y\right\Vert_{2} = \eta\left\Vert X-Y\right\Vert_{2}
\end{split}
\label{eta}\end{equation}
where $\eta=\frac{C}{\sqrt{n}}$ and $\Vert . \Vert_{2}$ is the Euclidean norm in the surrounding space $\mathbb{R}^{n}\supset S^{(n-1)}$.

Using this Lipschitz constant $\eta$ (Eqn.(\ref{eta})), Levy's lemma (Eqn.(\ref{levy})) now becomes,

\begin{equation}
    \text{Pr}\{ |f(X) - E[f(X)]| \geq \epsilon \} \leq \exp\left(-\frac{Kn^{2}\epsilon^{2}}{C^{2}}\right)
\label{levy1}\end{equation}

where $K$ and $C$ are some positive constants, for all $\epsilon \geq 0$.

Hence, we arrive at a Gaussian like functional upper bound (tighter than the exponential upper bound) which suggests that the function $f$ is very densely concentrated close to the expectation value $\mathbb{E}f$, closer than what the usual Levy's lemma suggests for points uniformly distributed on a n-dimensional hypersphere.

%This is illustrated in figures \ref{fig:eig_n} and \ref{fig:eig_e}, which is plotted from numerical calculations.

We could conclude that for any point picked at random in a high dimensional system, %shouldn't it be hypersphere?
the value of the fitness function will be concentrated around $\bar{f}=\mathbb{E}[f(X)]$ which is the mean of $f(X)$ taken over all values of $X$, for a given set of $\{a_{i}\}$ (Eqn. \ref{levy1}). Any random point $(x_{1},x_{2},\dots x_{n})$ represents a possible frequency distribution of the quasi-species. Modification of Levy's lemma, therefore shows that almost all quasi-species in higher dimensions have closely the same mean fitness.

\begin{figure}
    \centering
    \includegraphics[width=9.0cm, angle=0]{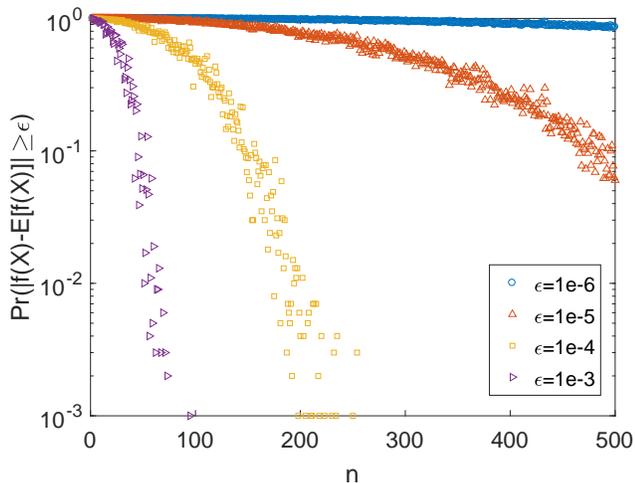}
    \caption{Probability of $f(X)$ to be in the $\epsilon$-neighbourhood of its mean $\mathbb{E}[f(X)]$ as a function of $n$, when replication and mutation rates are identically and independently picked from the exponential distribution (with unit mean) and then normalized. The probability has been plotted for different $\epsilon$ values on a semilog plot.}
    \label{fig:eig_n}
\end{figure}

\begin{figure}
    \centering
    \includegraphics[width=9.0cm, angle=0]{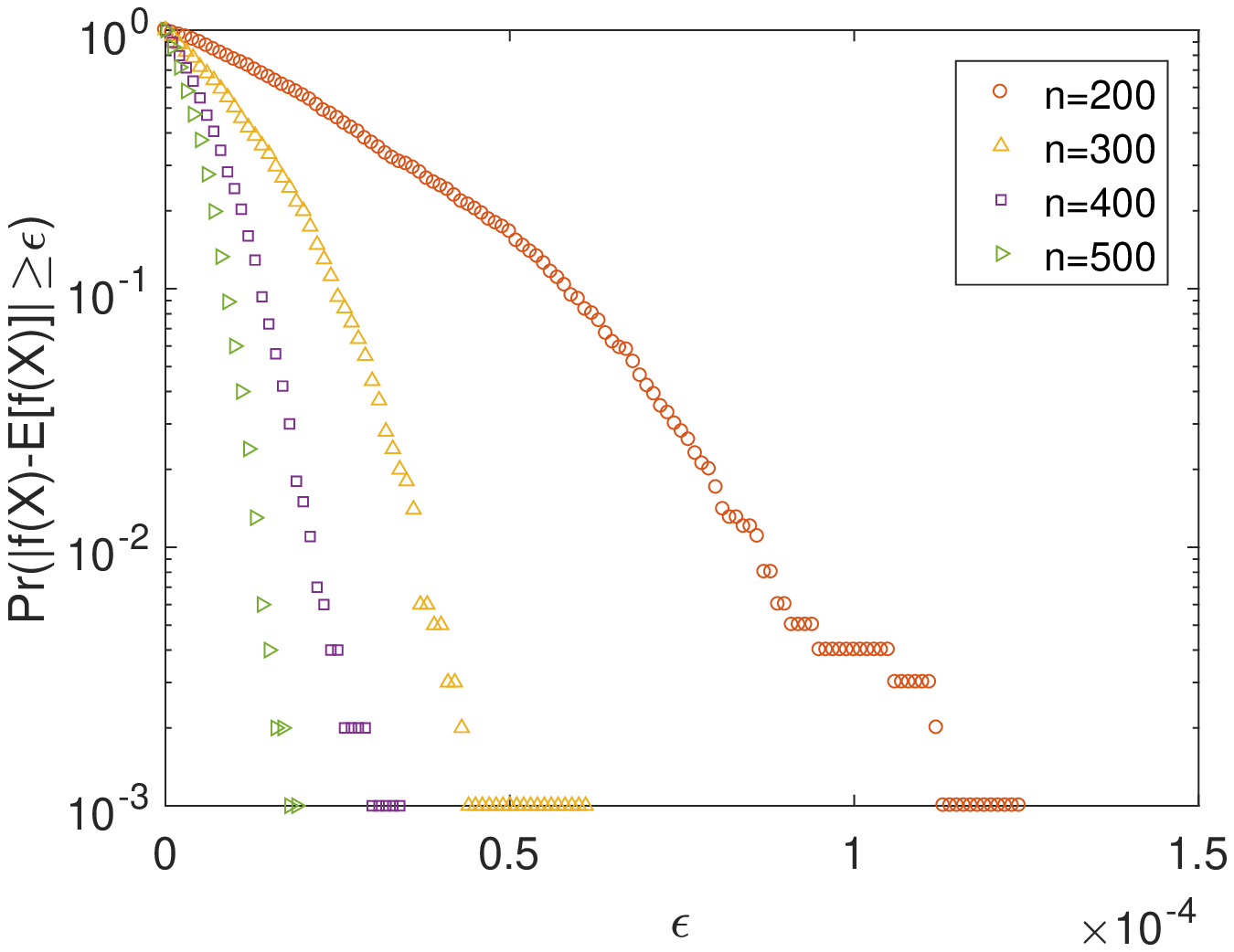}
    \caption{Probability of $f(X)$ to be in the $\epsilon$-neighbourhood of it's mean $\mathbb{E}[f(X)]$ as a function of $\epsilon$, when replication and mutation rates are identically and independently picked from the exponential distribution (with unit mean) and then normalized. The probability has been plotted for different $n$ values on a semilog plot.}
    \label{fig:eig_e}
\end{figure}

We have shown this result using numerical simulation too (figures \ref{fig:eig_n},\ref{fig:eig_e}). For a $n$-dimensional or a $n$-mutant quasi-species, we have assumed $n$ replication rates, $\{a_{i}\}$'s to be IID variables that are picked from the exponential distribution with unit mean ($f_{A}(a)=e^{-a}$) and then normalized. Similarly, we have repeated the same process with each of the $n$ rows of $Q_{ij}$ too, so that after normalization $\sum_{j=1}^{n}a_{j}=1$ and $\sum_{j=1}^{n}Q_{ij}=1$ for every $i$. Using the eigen value solvers, we have solved the eigenvalue equation (\ref{WX}) and obtained the eigenvector $X_{eq}=(x_{1},x_{2},\dots,x_{n})$ corresponding to the largest eigenvalue $\lambda_{max}$. This $\lambda_{max}=\sum_{j=1}^{n}a_{j}x_{j}$ (for normalized $\{a_{j}\}$'s and $\{x_{j}\}$'s) gives the fitness function $f(X)$. Now, this exercise is repeated for several times for different $Q$ matrices, all chosen similarly like before and the number of times $f(X)$ is in the $\epsilon$-neighbourhood of the mean $\mathbb{E}[f(X)]$ is calculated and normalized (to determine the probability). This is plotted for different values of $n$ and $\epsilon$ in figures (\ref{fig:eig_n},\ref{fig:eig_e}). We could observe that the quadratic nature of the exponential with respect to $n$ or $\epsilon$ is as given by Eqn. \ref{levy1}.

\subsection{Discussion - On the robustness of functional capabilities of quasi-species } 

 Any function, G, whose input parameters are the frequencies of the individual sequences, $X$, can be computed for the distribution given by the eigenvectors is concentrated closely about its average over the entire hypersphere. Therefore, if the functional behavior of quasi-species is given by such a function, its value is independent of the mutational matrix and the initial conditions. This also suggests robustness of functional behavior of quasi-species to perturbations in mutation rates and initial conditions. This is a significant result as one might expect that the workings of certain life processes, as described by quasi-species, require a certain degree of accuracy and robustness which we have shown is possible in high dimensional spaces.
 
Our analysis can be extend to the replicator-mutator equation \cite{hadeler, stadler, bomze}, which is used to describe the evolutionary dynamics of grammar and languages \cite{nowak}. 
 \begin{align}
\label{RM1}
 \frac{d X}{dt} = \hat{W} X - g(X). X
\end{align}
 The vector $X$ consists of the population densities of the individual sequences, 
  \begin{align}
\label{QS2}
 X = (x_1, x_2,..., x_n)
\end{align}
  
  The matrix $\hat{W}$ consists of individual replication rates, $\hat{a}_{i}(x_1, x_2,..., x_n), i = 1, 2, ..., n$, along with the mutation rates for transition between individual sequences, $i$ and $j$, given by $Q_{ij}$. The replication rates now are the functions of the frequency of the individual sequences.
  
\begin{align}
 \hat{W} = \begin{pmatrix}\hat{a}_1 Q_{11}&\hat{a}_2Q_{12}&...&\hat{a}_2Q_{1n} \\ \hat{a}_1 Q_{21}&\hat{a}_2Q_{22}&...&\hat{a}_2Q_{2n}\\...&...&...&...\\\hat{a}_1 Q_{n1}&\hat{a}_2Q_{n2}&...&\hat{a}_2Q_{nn} \end{pmatrix}_.
\end{align}  
  
  The total size of the population is a constant if we have 
  
  \begin{align}
\label{QS3}
g(X) = \sum_{i=1}^{n} \hat{a}_{i}x_{i} / \sum_{i}^{n} x_{i}
\end{align}.
 
We can map this solution on a hypersphere, analogous to the mapping of quasi-species. 
 The coordinates, $Y =(y_1, y_2, ..., y_n)$, of the hypersphere, are given by ($ {\sqrt{x_1}, \sqrt{x_2}, ..., \sqrt{x_n}}$).
The function, $f$, relevant to us, is $f=\sum_{i=1}^{n} \hat{a}_{i}x_{i} = \sum_{i=1}^{n} \hat{a}_{i}(y_1, y_2, ..., y_n)y_{i}^2$. If $\hat{a}_{i}(y_1, y_2, ..., y_n)y_{i}^2$ are bounded and Lipschitz for all $i$, then $f$ is Lipschitz and we can apply Levy's Lemma. Application of Levy's Lemma shows us that for any point picked at random on a high dimensional hypersphere, 
 the value of the function, $f$,  will be concentrated around $\bar{f}$ with high probability. For example, when describing the evolution of grammar,  the function $f$ is related to the \textit{grammatical coherence}, which quantifies the probability that a sentence said by one person is understood by other, will be robust to mutational rates and initial conditions \cite{nowakbook}. 
 
We can similarly show that any Lipschitz function, with input parameters given by the individual frequencies obtained from solving the replicator-mutator equation is concentrated closely about its average over the entire hypersphere. Therefore, if the functional capabilities of the system are described by a Lipschitz function with the individual frequencies as input parameters, we expect the value of the function to be concentrated about its average calculated over the hypersphere.

 \section{Conclusion}
 We have shown that the fitness of quasi-species is kinematical in nature, i.e. dependent on the system dimensions and individual selection rates and independent of the mutation dynamics and initial conditions.
  For almost all initial quasi-species distributions and mutation error probabilities, evolution leads to almost the same value of mean fitness %for fitness fitness 
  as defined by the largest eigenvalue of the mutation-selection matrix.
We have also shown how the functional capabilities of quasi-species is robust to mutational changes and fluctuations in the initial conditions. Our work is a consequence of application of ideas from high dimensional geometry to demonstrate the robustness of certain life processes and should be of use to explore the questions related to the origin of life.

\appendix

\section{Appendix-1: Probability Distribution of frequency distributions}

%Assume that we have IID variables $x_{1},x_{2},\dots,x_{n}$, each picked from an exponential distribution with mean $\lambda =1$. Let the normalized variables be defined as $u_{i}=\frac{x_{1}}{\sum_{i=1}^{n}x_{i}}$. We know that any $x=x_{i}$ is distributed as: $f_{X}(x)= e^{- x}$, by definition. Similarly, if $y=\sum_{j=1}^{n}x_{j}$ (where each $x_{i}$ is independent and exponentially distributed with the same mean $\lambda=1$), then we know that $y$ follows Erlang distribution: $f_{Y}(y)=y^{n-1}e^{-y}/(n-1)!$ (as given by the definition of Erlang distribution).

%Now, using variable transformation, we could determine the distribution of the variable $u=u_{i}=\frac{x_{i}}{y}=\frac{x_{i}}{\sum_{j=1}^{n}x_{j}}$ as:
%\begin{equation}
%\begin{split}
%    f_{U}(u)&= \frac{1}{(n-1)!}\int_{u}^{\infty}v^{n} e^{- uv}e^{-v}dv\\
%    &= \frac{1}{(n-1)!}\int_{u}^{\infty}v^{n}e^{-v(u+1)}dv\\
%    &= \frac{\Gamma(n+1,u(u+1))}{(n-1)!(u+1)^{n+1}}
%\end{split}
%\label{fu}\end{equation}
%where $\Gamma(s,x)$ is the upper incomplete gamma function. If we assume that $u\ll 1$, then $\Gamma(n+1,\lambda u(u+1))\approx \Gamma(n+1) =n!$. Similarly, the denominator could be approximated to $(1+u)^{n+1}\approx e^{u(n+1)}\approx e^{un}$ as $n \gg 1$. Hence, Eqn. \ref{fu} becomes,
%\[f_{U}(u)\approx ne^{-un}\],
%i.e. $u=u_{i}$ is exponentially distributed with mean $1/n$ satisfying $\sum_{i=1}^{n}u_{i}=1$

Assume that we have IID variables $x_{1},x_{2},\dots,x_{n}$, each picked from an exponential distribution with mean $\lambda =1$. Let the normalized variables be defined as $u_{i}=\frac{x_{1}}{\sum_{i=1}^{n}x_{i}}$. We know that any $x=x_{i}$ is distributed as: $f_{X}(x)= e^{- x}$, by definition. Similarly, if $y_{i}=\sum_{j=1,j\neq i}^{n}x_{j}$ (where each $x_{j}$ is independent and exponentially distributed with the same mean $\lambda=1$), then we know that $y=y_{i}$ follows Erlang distribution: $f_{Y}(y)=y^{n-2}e^{-y}/(n-2)!$ (as given by the definition of Erlang distribution for $n-1$ degrees of freedom).

Now, using change of variables technique, we could determine the distribution of the variable $u=u_{i}=\frac{x_{i}}{x_{i}+y_{i}}=\frac{x_{i}}{\sum_{j=1}^{n}x_{j}}$ as:
\begin{equation}
\begin{split}
    f_{U}(u)&= \frac{1}{(1-u)^{2}(n-2)!}\int_{0}^{\infty}v^{n-1} e^{-\frac{uv}{1-u}}e^{-v}dv\\
    &= \frac{1}{(1-u)^{2}(n-2)!}\int_{0}^{\infty}v^{n-1}e^{-\frac{v}{1-u}}dv\\
    &= \frac{\Gamma(n)(1-u)^{n-2}}{(n-2)!} = (n-1)(1-u)^{n-2}
\end{split}
\label{fu}\end{equation}
where $\Gamma(n)$ is the gamma function. For higher values of $n$, the last step could be approximated as $(n-1)(1-u)^{n-2}\approx (n-1)e^{-u(n-2)}\approx ne^{-un}$ as $n \gg 1$. Hence, Eqn. (\ref{fu}) becomes,
\[f_{U}(u)\approx ne^{-un}\]
i.e. $u=u_{i}$ is exponentially distributed with mean $1/n$ while satisfying $\sum_{i=1}^{n}u_{i}=1$

\section{Appendix-2: Modifying Levy's lemma}

From \textit{Appendix-1}, we know that each of the $x_{i}$ could be considered to be exponential IID random variables that satisfy $\sum_{i=1}^{n}x_{i}=1$. If we now consider the points $(y_{1},y_{2},\dots,y_{n}) = Y = \sqrt{X_{eq}}=(\sqrt{x_{1}},\sqrt{x_{2}},\dots,\sqrt{x_{n}})$ which would lie on a n-dimensional hypersphere $S^{(n-1)}$ with each of the coordinates $y_{i}$ distributed as $f_{Y}(y)\approx 2nye^{-ny^{2}}$ (as $f_{X}(x)\approx ne^{-nx}$), we would be able to extend Levy's lemma for frequency distributions. Also Ref. \cite{gerken} suggests that % new stuff
the functional value of $f(X)$ is $\epsilon$ away from its median value $M_{f}$, at most with probability given by twice the concentration function $\alpha_{\hat{X}}(\epsilon)$ taken over the entire domain $\hat{X}$. Here, $S^{(n-1)}$ is assumed as the domain of $X$, and so we have,

\begin{equation}
    \text{Pr}\{ |f(X) - M_{f}| \geq \epsilon \} \leq 2 \alpha_{S^{(n-1)}}(\epsilon)
\label{median_bound}\end{equation}

Actually, the concentration function $\alpha_{\hat{X}}(\epsilon)$, for a metric measure space $\hat{X}$ and for every $\epsilon >0$, could be defined as:

\[
    \alpha_{\hat{X}}(\epsilon) := \sup\{\mu(\hat{X} \setminus N_{\epsilon}(S)) | S \textit{ is measurable and } \mu(S)=\frac{1}{2}\}
\]
where $N_{\epsilon}(S)$ is the $\epsilon$-neighbourhood of $S$:
\[
    N_{\epsilon}(S) = \{x\in \hat{X} | \exists s\in S:d(s,x)<\epsilon\}
\]

To determine $\alpha_{\hat{X}}(\epsilon)$, we need to first define a spherical cap. A spherical cap $B(a,r)$ centered at point $a$ and radius $r$, is just a portion of a sphere cut off by a plane. By the Isoperimetric inequality for the sphere, we know that the measure on the unit sphere with the smallest border or $\epsilon$-expansion, is the cap $B(a,r)$, as $B(a,r)\subset S^{(n-1)}\subset \mathbb{R}^{n}$ solves the isoperimetric problem for the sphere. This result could similarly be extended to higher dimensions too.

Since the uniformly distributed points on a n-dimensional hypersphere and $Y$ share similar axial symmetry on the hypersphere, we could consider the same spherical cap $B(a,r) \subset S^{(n-1)}$ around one of the polar points $a\in S^{(n-1)}$, with respect to $Y$ and with radius $r$ (given by angular norm) for calculating the concentration function $\alpha_{S^{(n-1)}}(\epsilon)$ here.
\begin{equation}
    \alpha_{S^{(n-1)}}(\epsilon)=1-\mu(B(a,\frac{\pi}{2}+\epsilon)) = 1-A(\phi)
\label{alpball}\end{equation}

In order to determine $A(\phi)$, we need to know how the points $Y$ are distributed over $S^{(n-1)}$. Actually, since we are only concerned about the spherical cap $B(a,\frac{\pi}{2}+\epsilon)\subset S^{(n-1)}$, we would only require the distribution of $\phi_{1}$, where $\phi_{1}$ is the $1^{st}$ or the principal angular coordinate. For an n-dimensional unit hypersphere $S^{(n-1)}$, $y_{1}=\cos(\phi_{1})$,
\begin{equation}
    f_{\Phi_{1}}(\phi_{1})=\frac{n}{2}|\sin(2\phi_{1})|\exp(-n\cos^{2}(\phi_{1}))
\label{fphi1}\end{equation}

if we allow $y_{1}$ to take negative values too. Using the volume element for $S^{(n-1)}$, we could calculate $A(\phi)$,

\begin{equation}
    A(\phi) = s_{n-1}^{-1}\int_{0}^{\phi}f_{\Phi_{1}'}(\phi_{1}')\sin^{n-2}(\phi_{1}')d\phi_{1}'
\label{aphi}\end{equation}

where $s_{n-1}=\int_{0}^{\pi}f_{\Phi_{1}'}(\phi_{1}')\sin^{n-2}(\phi_{1}')d\phi_{1}'$.

Combining Eqns. \ref{alpball}, \ref{fphi1} and \ref{aphi}, we get,

\begin{equation}
\begin{split}
    \alpha_{S^{(n-1)}}(\epsilon) &= 1 - s_{n-1}^{-1}\int_{0}^{\frac{\pi}{2}+\epsilon}f_{\Phi_{1}'}(\phi_{1}')\sin^{n-2}(\phi_{1}')d\phi_{1}'\\
    &= s_{n-1}^{-1}\int_{\frac{\pi}{2}+\epsilon}^{\pi}f_{\Phi_{1}'}(\phi_{1}')\sin^{n-2}(\phi_{1}')d\phi_{1}'\\
    &= s_{n-1}^{-1}\int_{\frac{\pi}{2}+\epsilon}^{\pi}\frac{n}{2}|\sin(2\phi_{1}')|\times\\
    &\exp\left(-n\cos^{2}(\phi_{1}')\right)\sin^{n-2}(\phi_{1}')d\phi_{1}'\\
    &= s_{n-1}^{-1}\int_{0}^{\cos\epsilon}n\exp\left(n(t^{2}-1)\right)t^{n-1}dt\\
    &= \frac{n}{2s_{n-1}}e^{-n}(-n)^{-n/2}\left((n/2)-1\right)!\times\\
    &\left[1-e^{n\cos^{2}\epsilon} \sum_{k=0}^{(n/2)-1}\frac{(-n\cos^{2}\epsilon)^{k}}{k!} \right]
\end{split}
\label{alpderive}\end{equation}

Similarly, integrating the expression for $s_{n-1}^{-1}$ and combining it with Eqn. \ref{alpderive}, we get (after cancelling some terms),

\begin{equation}
\begin{split}
    \alpha_{S^{(n-1)}}(\epsilon) &= \frac{\left[1-e^{n\cos^{2}\epsilon} \sum_{k=0}^{\frac{n}{2}-1}\frac{(-n\cos^{2}\epsilon)^{k}}{k!} \right]}{2\left[1-e^{n\epsilon} \sum_{k=0}^{\frac{n}{2}-1}\frac{(-n)^{k}}{k!} \right]}\\
    &= \frac{\left(e^{-n\sin^{2}\epsilon}\right)}{2}\frac{\sum_{k=\frac{n}{2}}^{\infty}\frac{(-n\cos^{2}\epsilon)^{k}}{k!}}{\sum_{k=\frac{n}{2}}^{\infty}\frac{(-n)^{k}}{k!}}\\
    &\leq \frac{\left(e^{-n\sin^{2}\epsilon}\right)}{2}\cos^{n}\epsilon \leq \frac{1}{2}\left(e^{-n\sin^{2}\epsilon-\frac{n\epsilon^{2}}{2}}\right)\\
    &\leq \frac{1}{2}\left(e^{-\frac{n\epsilon^{2}}{2}}\right)
\end{split}
\label{alpderive1}\end{equation}

Since $\alpha_{X}^{e}(\epsilon)\leq \alpha_{X}^{a}(\epsilon)$, (Ref. \cite{gerken}) where $\alpha_{X}^{e}(\epsilon)$ is the Euclidean concentration function and $\alpha_{X}^{a}(\epsilon)$ is the angular concentration function, we could combine Eqn. (\ref{alpderive1}) and Eqn. (\ref{median_bound}) to get,

\[
    \text{Pr}\{ |f(X) - M_{f}| \geq \epsilon \} \leq \exp\left(-n\epsilon^{2}/2\right)
\]

Modifying $\epsilon \rightarrow \frac{\epsilon}{\eta}$,

\[
    \text{Pr}\{ |f(X) - M_{f}| \geq \epsilon \} \leq \exp\left(-\frac{n\epsilon^{2}}{2\eta^{2}}\right)
\]

As mentioned in Ref. \cite{gerken}, since median $M_{f}$ and expectation value $\mathbb{E}f$ are not the same, we could make some modifications to the factors of the exponential function to change the expression from median based to expectation value based.

\begin{equation}
    \text{Pr}\{ |f(X) - \mathbb{E}f| \geq \epsilon \} \leq \exp\left(-\frac{Kn\epsilon^{2}}{\eta^{2}}\right)
\label{levymod}\end{equation}

where $K$ is some positive constant.% Perhaps $K=\frac{3}{18\pi^{3}}$

\section{Appendix-3: Determining Lipschitz constant}

In order to determine the Lipschitz constant of the function $f(X)$, we could start by determining how change in each of the coordinates changes the functional value.
We see that,

\begin{equation}
\begin{split}
    |f - \hat{f}| &\leq a_{i}| x_{i}-\hat{x}_{i}|\\
    &\leq a_{max}| x_{i}-\hat{x}_{i}|
\end{split}
\label{fbound1}\end{equation}
where $f$ and $\hat{f}$ are functional values when only one of the coordinates, namely $x_{i}$ is changed to $\hat{x}_{i}$.

That makes $f$, Lipschitz continuous along the $x_{i}$ coordinate, with Lipschitz constant $a_{max}$. Let, $X,Y \in D=[0,1]^{n}/[0,\delta]^{n}$, $X=(x_{1},x_{2},\dots,x_{n})$ and $Y=(y_{1},y_{2},\dots,y_{n})$, then we know,

\begin{equation}
\begin{split}
    f(X)-f(Y)=f(x_{1},x_{2},\dots,x_{n})-f(y_{1},y_{2},\dots,y_{n})\\ 
    = f(x_{1},x_{2},\dots,x_{n-1},x_{n})- f(x_{1},x_{2},\dots,x_{n-1},y_{n}) +\\  f(x_{1},x_{2},\dots,x_{n-1},y_{n}) - f(x_{1},x_{2},\dots,y_{n-1},y_{n}) +\\  f(x_{1},x_{2},\dots,y_{n-1},y_{n}) - \hdots \hdots \hdots \hdots \hdots \hdots \hdots +\\
    f(x_{1},y_{2},\dots,y_{n-1},y_{n}) -f(y_{1},y_{2},\dots,y_{n-1},y_{n})\\
\end{split}
\label{fxy}\end{equation}

Using, triangle inequality, we would then get,

\begin{equation}
\begin{split}
    |f(X)-f(Y)| \leq |f(x_{1},x_{2},\dots,x_{n})- f(x_{1},x_{2},\dots,y_{n})|+\\
    |f(x_{1},\dots,x_{n-1},y_{n}) - f(x_{1},\dots,y_{n-1},y_{n})|+\\
    \dots +|f(x_{1},y_{2},\dots,y_{n}) -f(y_{1},y_{2},\dots,y_{n})|\\
\end{split}
\label{flipschitz}\end{equation}

and from Eqn. (\ref{fbound1}),

\begin{equation}
    \leq a_{max}\left(\sum_{i=1}^{n}|x_{i}-y_{i}|\right)\\
\label{flipschitz1}\end{equation}

Now, using Cauchy-Schwarz inequality, we get,

\begin{equation}
\begin{split}
    \leq \sqrt{n}a_{max}\left(\sum_{i=1}^{n}|x_{i}-y_{i}|^{2}\right)^{1/2}\\
    = \sqrt{n}a_{max}\left\Vert X-Y\right\Vert_{2} = \eta\left\Vert X-Y\right\Vert_{2}
\end{split}
\label{flipschitz2}\end{equation}

where $\eta = \sqrt{n}a_{max}$. Hence, function $f$ is Lipschitz continuous with Lipschitz constant, $\eta$. Basically, the function becomes Lipschitz continuous as long as we make sure there is no singularity in the domain. %For a stronger bound, we'd used the assumption that, %$\text{Range}(\{a_{i}\}) = |a_{M}-a_{m}| \leq \frac{C}{n}$ with very high probability. This means (from %Eqn. (\ref{flipschitz})),
%\section{Appendix-4: Further numerical evidence of }

\begin{figure}
    \centering
    \includegraphics[width=9.0cm, angle=0]{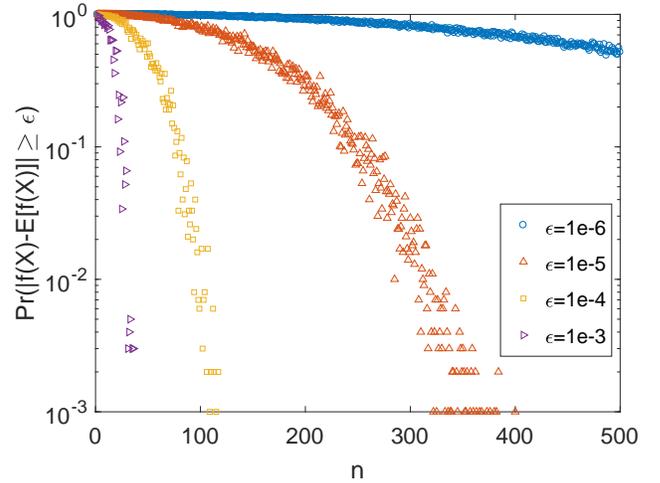}
    \caption{Probability of $f(X)$ to be in the $\epsilon$-neighbourhood of its mean $\mathbb{E}[f(X)]$ as a function of $n$, when replication and mutation rates are identically and independently picked from the uniform distribution $[0,1]$ and then normalized. The probability has been plotted for different $\epsilon$ values on a semilog plot.}
    \label{fig:eig_n2}
\end{figure}

\begin{figure}
    \centering
    \includegraphics[width=9.0cm, angle=0]{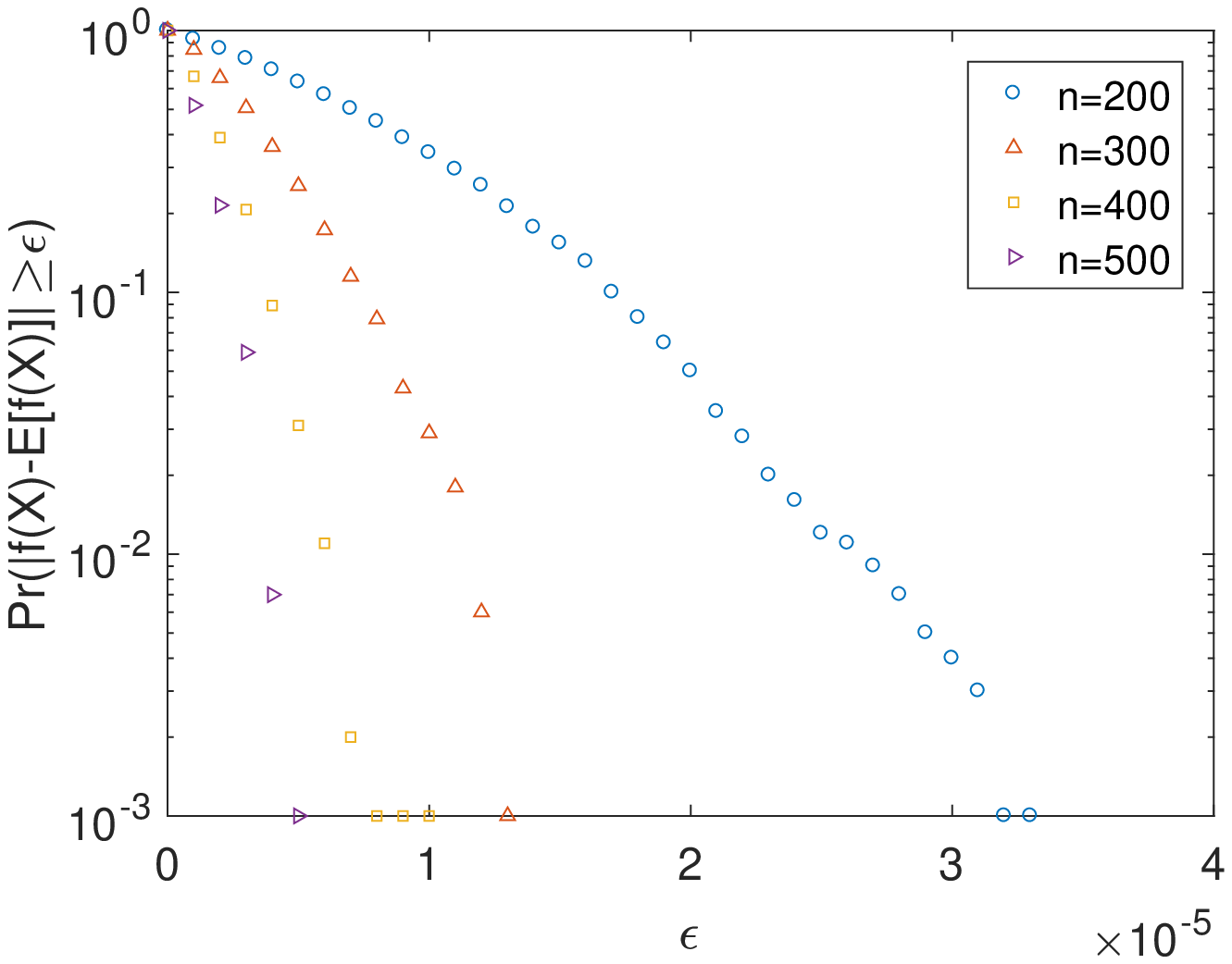}
    \caption{Probability of $f(X)$ to be in the $\epsilon$-neighbourhood of it's mean $\mathbb{E}[f(X)]$ as a function of $\epsilon$, when replication and mutation rates are identically and independently picked from the uniform distribution $[0,1]$ and then normalized. The probability has been plotted for different $n$ values on a semilog plot.}
    \label{fig:eig_e2}
\end{figure}

\section{Supplementary Material}

The previous analysis assumed we have IID variables $x_{1}, x_{2},...., x_{n}$, each
picked from an exponential distribution with mean $\lambda = 1$. We then found the appropriate distribution for the points $(y_{1},y_{2},\dots,y_{n}) = Y = \sqrt{X_{eq}}=(\sqrt{x_{1}},\sqrt{x_{2}},\dots,\sqrt{x_{n}})$ which would lie on an n-dimensional hypersphere $S^{(n-1)}$ with each of the coordinates $y_{i}$ distributed as $f_{Y}(y)\approx 2nye^{-ny^{2}}$ (as $f_{X}(x)\approx ne^{-nx}$). We then extended Levy's lemma to find the concentration of measure properties associated with this distribution.

In this section, we extend our arguments to general distributions for the coordinates for the points $(y_{1},y_{2},\dots,y_{n}) = Y$. The only assumptions we make is that the probability distribution for $Y$ is a Lipschitz function, $\mathcal{P}(Y)$, with the Lipschitz constant $\eta$. We can consider the probability distribution itself to be a function on the sphere
whose inputs are points picked at random from a uniform distribution over the sphere. The value of the function gives the value of the probability density as a function of the coordinates.

Then, applying the Levy's Lemma to this function, 

\begin{equation}
    \text{Pr}\{ |\mathcal{P(Y)} - \mathbb{E}\mathcal{P}| \geq \epsilon' \} \leq \exp\left(-\frac{K'n\epsilon'^{2}}{\eta'^{2}}\right)
\label{levymod0}\end{equation}

where $K'$ is some positive constant, which suggests that the function $\mathcal{P}$ is very densely concentrated close
to the expectation value $\mathbb{E}\mathcal{P}$. 

\begin{eqnarray}
\label{Eq:Ereal}
\mathbb{E}{\mathcal{P}} &=&\frac{  \int_{S^{n-1}} \mathcal{P}(y_1, y_2,..., y_n) d \mu}{ \int_{S^{n-1}} d \mu} = \mathcal{C}\nonumber \\
\end{eqnarray}

Here, $d \mu$ is the surface area of a differential patch on a hypersphere - the probability measure for picking uniformly distributed points. The integral in the numerator above equals one (as $\mathcal{P}$ is a normalized probability distribution), we can conclude that  $\mathbb{E}{\mathcal{P}}$ is a constant, $\mathcal{C}$, for all probability distributions. In the units we are working, where $\mu(S^{n-1}) = 1$, $\mathcal{C}$ is equal to one.
Therefore, 
the value of  $\mathcal{P}$ is very densely concentrated about $\mathcal{C}$ and is equal to that of the uniform distribution.  
Let a point, $X$, be chosen at random with respect to $\mathcal{P}$. The new concentration function associated with $\mathcal{P}(X)$ is given by evaluating the probability that $X$ lies 
in the region $S^{n-1}/B(a,\frac{\pi}{2}+\epsilon)$. 
And the Levy' s lemma reads, 
  
   % $\text{Pr}\{ |f(X) - \mathbb{E}f| \geq \epsilon \} $
%\times  \exp\left(-\frac{Kn\epsilon^{2}}{\eta^{2}}\right).

\begin{widetext}
\begin{equation}
    \text{Pr}\{ |f(X) - \mathbb{E}f| \geq \epsilon \} \leq 2 \bigg( \int_{ S^{n-1}\setminus B(a,\frac{\pi}{2}+\epsilon)} \mathcal{P}(X)  d \mu \bigg) \leq 2 \bigg( \int_{ S^{n-1}\setminus B(a,\frac{\pi}{2}+\epsilon)} (1+\epsilon')  \exp\left(-\frac{K'n\epsilon'^{2}}{\eta'^{2}}\right) d \mu \bigg)
\label{levymod1}
\end{equation}
\end{widetext}
Since,  $\mu(S^{n-1}\setminus B(a,\frac{\pi}{2}+\epsilon)) =  \alpha_{S^{(n-1)}}(\epsilon)$, 
the right most integral in the above inequality can be evaluated to be equal to 

%\begin{widetext}
\begin{equation}
    \text{Pr}\{ |f(X) - \mathbb{E}f| \geq \epsilon \}  \leq 2  (1+\epsilon')  \exp\left(-\frac{K'n\epsilon'^{2}}{\eta'^{2}}\right) \alpha_{S^{(n-1)}}(\epsilon)
\label{levymod2}
\end{equation}
%\end{widetext}

The terms involving $\epsilon'$,  $(1+\epsilon')  \exp\left(-\frac{K'n\epsilon'^{2}}{\eta'^{2}}\right)$ can be bounded by a constant, $\lambda$, depending on values of  $K'$ and $\eta'$.
Putting back the value of $\alpha_{S^{(n-1)}}(\epsilon)$ evaluated in \textit{Appendix-2}, we get 
\begin{equation}
    \text{Pr}\{ |f(X) - \mathbb{E}f| \geq \epsilon \}  \leq   \lambda  \exp\left(-\frac{Kn\epsilon^{2}}{\eta^{2}}\right)
\label{levymod3}
\end{equation}

Thus, as long as points are taken from a  probability distribution that is a Lipschitz function, $\mathcal{P}(Y)$, with the Lipschitz constant $\eta$, our results will hold.
Figure \ref{fig:eig_n2} and Figure \ref{fig:eig_e2} are further numerical evidence of our finding. These figures are plotted from numerical simulations exactly similar to the one performed for generating the figures \ref{fig:eig_n} and \ref{fig:eig_e}, except that now we have assumed the $n$ replication rates, $\{a_{i}\}$'s to be IID variables that are picked from the uniform distribution $[0,1]$ (instead of an exponential) and then normalized it. Similarly, we have repeated the same process with each of the $n$ rows of $Q_{ij}$ too, so that after normalization, like earlier, $\sum_{j=1}^{n}a_{j}=1$ and $\sum_{j=1}^{n}Q_{ij}=1$ for every $i$. The subsequent procedures are repeated exactly like before and the simulation is repeated for different choices of $Q$ and the probability that $f(X)$ is in the $\epsilon$-neighbourhood is plotted for different values of $n$ and $\epsilon$ in figures (\ref{fig:eig_n2},\ref{fig:eig_e2}). We could again observe the quadratic nature of the exponential with respect to $n$ or $\epsilon$ as given by Eqn. \ref{levy1}, which reinforces our results.

%1) make the references uniform - done

%2) a little bit more description of the spherical cap...levy's lemma - done

%3) spell check - done

%4) details of fig 3 and 4. how the matrices were chosen. also the captions of fig 3 and 4 - done

%\begin{equation}
%\begin{split}
%    |f(X)-f(Y)| &\leq \frac{\sqrt{n}}{\delta}|a_{M}-a_{m}|\left\Vert X-Y\right\Vert\\
%    &\leq \frac{C}{\delta\sqrt{n}}\left\Vert X-Y\right\Vert\\
%    &\leq \frac{C}{\delta}\left\Vert X-Y\right\Vert = \eta_{2} \left\Vert X-Y\right\Vert
%\end{split}
%\label{lipschitz}\end{equation}

%Hence, function $f$ is Lipschitz continuous with Lipschitz constant, $\eta$. Basically, the function becomes Lipschitz continuous once we make sure there is no singularity.

\begin{acknowledgements}
VM acknowledges discussions with Karen Page and Michael Doebeli. VM acknowledges IIT Madras for financial support. %Unacknowledgement: This manuscript was delayed due to the inconvenience caused by stray dogs and their lovers in the IIT campus.
\end{acknowledgements}

\end{document}